\documentclass[12pt]{article}
\begin{document}
\baselineskip=16pt
\begin{titlepage}
\setcounter{page}{0}
\begin{center}

\vspace{0.5cm} {\Large \bf Cosmological Scaling Solutions and
Cross-coupling Exponential Potential }\\
\vspace{10mm}
Zong-Kuan Guo\footnote{e-mail address: guozk@itp.ac.cn}$^{a}$,
Yun-Song Piao\footnote{e-mail address: yspiao@itp.ac.cn}$^{a}$,
Rong-Gen Cai$^{a}$
and
Yuan-Zhong Zhang$^{b,a}$ \\
\vspace{6mm} {\footnotesize{\it
  $^a$Institute of Theoretical Physics, Chinese Academy of Sciences,
      P.O. Box 2735, Beijing 100080, China\\
  $^b$CCAST (World Lab.), P.O. Box 8730, Beijing 100080\\}}

\vspace*{5mm} \normalsize
\smallskip
\medskip
\smallskip
\end{center}
\vskip0.6in 
\centerline{\large\bf Abstract}
{We present a phase-space analysis of cosmology containing multiple scalar
fields with a positive or negative cross-coupling exponential potential.
We show that there exist power-law kinetic-potential-scaling solutions for
a sufficiently flat positive potential or for a steep negative potential.
The former is the unique late-time attractor, but it is difficult to yield
assisted inflation. The later is never stable in an expanding universe.
Moreover, for a steep negative potential there exists a kinetic-dominated
regime in which each solution is a late-time attractor. In the presence of
ordinary matter these scaling solutions with a negative cross-coupling
potential are found unstable. We briefly discuss the physical consequences
of these results.}

\vspace*{2mm}
\end{titlepage}

\section{Introduction}

Scalar field cosmological models are of great importance in modern 
cosmology. The dark energy is attributed to the dynamics of a 
scalar field, which convincingly realizes the goal of explaining 
current accelerating expansion of universe generically using only attractor 
solutions~\cite{CD}. Furthermore a scalar field can drive an
accelerated expansion and thus provides possible models for cosmological 
inflation in the early universe~\cite{GU}. In particular, there 
have been a number of studies of spatially homogeneous scalar 
field cosmological models with an exponential potential. They are 
already known to have interesting properties; for example, if one 
has a universe containing a perfect fluid and such a scalar field, 
then for a wide range of parameters the scalar field mimics the
perfect fluid, adopting its equation of state~\cite{CL}. These
scaling solutions are attractors at late times~\cite{ST}. The inflation
models and other cosmological consequences of multiple scalar fields
have also been considered~\cite{LM, PCZ}.

The scale-invariant form makes the exponential potential 
particularly simple to study analytically. There are well-known 
exact solutions corresponding to power-law solutions for the 
cosmological scale factor $a\propto t^p$ in a spatially flat 
Friedmann-Robertson-Walker (FRW) model~\cite{FLM}.
More generally the coupled 
Einstein-Klein-Gordon equations for a single field can be reduced 
to a one-dimensional system which makes it particularly suitable for
a qualitative analysis~\cite{JJH,CLW}. Recently, adopting a system 
of dimensionless dynamical variables~\cite{EW}, 
the cosmological scaling solutions with positive or negative 
exponentials have been studied~\cite{HW}. In general there are many 
scalar fields with exponential potentials in supergravity,
superstring and the generalized Einstein theories, thus multiple
potentials may be more interesting. In the previous paper~\cite{GP},
We studied the stability of cosmological scaling
solutions in an expanding universe model with multiple scalar fields with
positive or negative exponential potentials.
A phase-space analysis of the spatially flat FRW models shows that there
exist cosmological scaling solutions which are the unique late-time
attractors and successful inflationary solutions driven by multiple
scalar fields with a wide range of each potential slope parameter $\lambda$.
It is assumed that there is no direct coupling between potentials. 
Multiple cross-coupling exponential potentials arise in many occasions, for
instance, from compactifications of vacuum Einstein gravity 
on product spaces~\cite{CHN}. Indeed they are
a natural outcome of the compactification of higher dimensional theories
down to 3+1 dimensions. With this in mind it is worth investigating such
potential in a bit more detail.

In this paper, we first study a system of dimensionless dynamical
variables of multiple scalar
fields with a positive or negative cross-coupling exponential potential. We
obtain the scaling solutions and analyze their stability. There still exist
cosmological scaling solutions which are the unique late-time attractors.
In this model we then introduce a barotropic fluid to the system.
We discuss the physical consequences of these results.

\section{Cross-coupling Exponential Potential}

We consider $n$ scalar fields $\phi _i$ with a cross-coupling potential
\begin{equation} \label{CCP}
V=V_0\exp \left(-\sum_{i=1}^{n}\lambda _i \kappa \phi _i\right), 
\end{equation} 
where $\kappa ^2\equiv 8 \pi G_N$ is the gravitational coupling and
$\lambda _i$ are dimensionless constants characterising the slope of the
potential. Further we assume all $\lambda _i \ge 0$ since we can make
them positive through $\phi_i \to -\phi_i$ if some of them are negative. 
The evolution equation of each scalar field for a spatially
flat FRW model with Hubble parameter $H$ is
\begin{equation} \label{EE} 
\ddot{\phi _i}+3H\dot{\phi _i}-\lambda _i \kappa V=0,
\end{equation} 
subject to the Friedmann constraint 
\begin{equation} \label{CE}
H^2=\frac{\kappa ^2}{3}\left(\sum_{i=1}^{n}\frac{1}{2}\dot{\phi}_i^2
 +V\right).
\end{equation} 
Defining $(n+1)$ dimensionless variables 
\begin{equation} 
x_i=\frac{\kappa \dot{\phi _i}}{\sqrt{6}H}, \qquad 
y=\frac{\kappa \sqrt{|V|}}{\sqrt{3}H}, 
\end{equation} 
the evolution equations (\ref{EE}) can be written as an autonomous system: 
\begin{eqnarray} \label{AS1} 
x'_i &=& -3x_i\left(1-\sum_{j=1}^{n}x_j^2\right)
 \pm \lambda _i\sqrt{\frac{3}{2}}y^2, \\  
y'&=& y\sum_{j=1}^{n}\left(3x_j^2-\lambda _j\sqrt{\frac{3}{2}}x_j\right), 
\label{AS2}
\end{eqnarray}
where a prime denotes a derivative with respect to the logarithm of the
scalar factor, $N \equiv \ln a$, and the constraint equation (\ref{CE})
becomes 
\begin{equation} \label{CEE} 
\sum_{i=1}^{n}x_i^2 \pm y^2=1. 
\end{equation} 
Throughout this paper
we will use upper/lower signs to denote the two distinct cases
of $\pm V_0>0$. $x_i^2$ measures the contribution to the expansion due to
the field's kinetic energy density, while $\pm y^2$ represents the
contribution of the potential energy. We will restrict our discussion of
the existence and stability of critical points to expanding universes
with $H>0$, i.e., $y \ge 0$. Critical points correspond
to fixed points where $x'_i=0$ and $y'=0$, and there are self-similar
solutions with
\begin{equation} 
\frac{\dot{H}}{H^2}=-3\sum_{i=1}^{n}x_i^2. 
\end{equation} 
This corresponds to an expanding universe with a scale factor $a(t)$ given
by $a\propto t^p$, where 
\begin{equation}
p=\frac{1}{3\sum_{i=1}^{n}x_i^2}\,.
\end{equation}
The system (\ref{AS1}) and (\ref{AS2}) has at most one $n$-dimensional 
sphere $S$ embedded in $(n+1)$-dimensional phase-space corresponding to 
kinetic-dominated solutions, and a fixed point $A$, which is a
kinetic-potential-scaling solution listed in Table 1.

In order to study the stability of the critical points,
using the Friedmann constraint equation (\ref{CEE}) we first reduce
Eqs.(\ref{AS1}) and (\ref{AS2}) to $n$ independent equations
\begin{equation} \label{IE}
x'_i = \left(\lambda _i\sqrt{\frac{3}{2}}-3x_i\right)
\left(1-\sum_{j=1}^{n}x_j^2\right).
\end{equation}
Substituting linear perturbations $x_i \to x_i+\delta x_i$ about the
critical points into Eqs.(\ref{IE}), to first-order in the perturbations,
gives equations of motion
\begin{equation} \label{PE}
\delta x'_i = -2\left(\lambda _i\sqrt{\frac{3}{2}}-3x_i\right)
\sum_{j=1}^{n}(x_j \delta x_j)-
3\left(1-\sum_{j=1}^{n}x_j^2\right)\delta x_i\,,
\end{equation}
which yield $n$ eigenvalues $m_i$. Stability requires the real part of all
eigenvalues being negative.

$S$: $\sum_{i=1}^{n}x_i^2=1$, $y=0$.
These kinetic-dominated solutions always exist for any form of the
potential, which are equivalent to stiff-fluid dominated evolution with
$a\propto t^{1/3}$ irrespective of the nature of the potential.
Then Eqs.(\ref{PE}) become 
\begin{displaymath}
\delta x'_i = -2\left(\lambda _i\sqrt{\frac{3}{2}}-3x_i\right)
\sum_{j=1}^{n}(x_j \delta x_j),
\end{displaymath}
which yield $n$ eigenvalues: one of them, say $m_1$, does not vanish,
$m_1=-\sqrt{6}(\sum_{i=1}^{n}\lambda_i x_i-\sqrt{6})$; the remains
of them vanish. Thus the solutions are marginally
stable for $\sum_{i=1}^{n}(\lambda_i x_i)>\sqrt{6}$.
For the special case $\lambda_i=\lambda$, using the constraint equation
(\ref{CEE}) we find $\sqrt{6}/(n\lambda) < \sum_{i=1}^{n}x_i/n \le
(\sum_{i=1}^{n}x_i^2/n)^{1/2}=1/\sqrt{n}$. That is, if each scalar field
has an identical-slope potential, there exist stable points only for
$\lambda ^2 > 6/n$.

$A$: $x_i=\frac{\lambda _i}{\sqrt{6}}$,
$y=\sqrt{\pm(1-\frac{1}{6}\sum_{i=1}^{n} \lambda_i^2)}$.
The potential-kinetic-scaling solution exists for sufficiently flat
$\sum_{i=1}^{n}\lambda _i^2 < 6$ positive potentials or steep
$\sum_{i=1}^{n}\lambda _i^2 > 6$ negative potentials.
The power-law exponent, $p=\frac{2}{\sum_{i=1}^{n} \lambda_i^2}$,
depends on parameter $\lambda_i$. From Eqs.(\ref{PE}) we find
the eigenvalues 
\begin{displaymath}
m_i=-\frac{1}{2}\left(6-\sum_{j=1}^{n}\lambda _j^2\right).
\end{displaymath}
Thus the scaling solution is always stable when this point exists for a
positive potential, which corresponds to the power-law inflation in an
expanding universe when $\sum_{i=1}^n \lambda_i^2 < 2$.
However, this solution is unstable for a negative potential.

The different regions of $\lambda_i$ parameter space lead to different 
qualitative evolution. As an example we consider the cosmologies containing
$n$ scalar fields with the cross-coupling potential $\lambda_i=\lambda$.
For the sufficiently flat ($\lambda ^2<6/n$) positive potential, these
kinetic-dominated
solutions are unstable and the kinetic-potential-scaling solution is the
stable late-time attractor. Hence generic solutions start in the former
and approach the later at late times. For the steep ($\lambda ^2 > 6/n$)
positive potential, there exists a stable kinetic-dominated regime, in which
each points are the late-time attractors. Hence generic solutions start in
kinetic-dominated solution and approach the stable regime. For the flat
sufficiently
($\lambda ^2 < 6/n$) negative potential, only these kinetic-dominated
solutions exist which are unstable scaling solutions. For the steep 
($\lambda ^2 > 6/n$) negative potential, the kinetic-potential-scaling
solution is unstable and there exists a stable kinetic-dominated regime.
Hence generic solutions start in a kinetic-dominated regime or the
kinetic-potential-scaling solution and approach the stable kinetic-dominated
regime at late times.

\begin{table}

\begin{tabular}{||c|c|c|c|c||} \hline \hline 
Label & $x_i$ & $y$ & Existence & Stability \\ \hline
S & $\sum_{i=1}^{n}x_i^2=1$ & 0 & all $\lambda_i$ &
stable $\sum_{i=1}^{n}(\lambda_i x_i) > \sqrt{6}$ \\ \hline
A$_+$ & $\frac{\lambda_i}{\sqrt{6}}$ & 
$\sqrt{ (1-\sum_{i=1}^{n}\frac{\lambda _i^2}{6})}$ &
$\sum_{i=1}^{n} \lambda _i^2 < 6$, $V>0$ & stable  \\ \hline
A$_-$ & $\frac{\lambda_i}{\sqrt{6}}$ &
$\sqrt{ (\sum_{i=1}^{n}\frac{\lambda _i^2}{6}-1)}$ &
$\sum_{i=1}^{n} \lambda _i^2 > 6$, $V<0$ & unstable  \\
\hline \hline
\end{tabular}

\caption{The properties of the critical points in a spatially flat FRW
universe containing $n$ scalar fields with the cross-coupling exponential
potential.}

\end{table}

\section{Plus a Barotropic Fluid}

We now consider multiple scalar fields with the cross-coupling potential
(\ref{CCP}) evolving in a spatially flat FRW universe containing a fluid
with barotropic equation of state $P_\gamma=(\gamma-1)\rho_\gamma$, where
$\gamma$ is a constant, $0 < \gamma \le 2$, such as radiation ($\gamma=4/3$)
or dust ($\gamma=1$). The evolution equation for the barotropic fluid is
\begin{equation} \label{BF}
\dot{\rho_ \gamma}=-3H(\rho_\gamma+P_\gamma),
\end{equation}
subject to the Fridemann constraint
\begin{equation}
H^2=\frac{\kappa ^2}{3}\left(\sum_{i=1}^{n}\frac{1}{2}\dot{\phi}_i^2
 +V+\rho_\gamma \right).
\end{equation}
We define another dimensionless variable
$z \equiv \frac{\kappa \sqrt{\rho_\gamma}}{\sqrt{3}H}$. The evolution
equations (\ref{EE}) and (\ref{BF}) can then be written as an autonomous
system:
\begin{eqnarray} \label{AAS1} 
x'_i &=& -3x_i\left(1-\sum_{j=1}^{n}x_j^2-\frac{\gamma}{2}z^2\right)
 \pm \lambda _i\sqrt{\frac{3}{2}}y^2, \\  
y' &=& y \left(3\sum_{i=1}^{n}x_i^2+\frac{3\gamma}{2}z^2
 -\sqrt{\frac{3}{2}}\sum_{i=1}^{n}\lambda_i x_i\right), \\
z' &=& \frac{3}{2}z\left(-\gamma+2\sum_{i=1}^{n}x_i^2+\gamma z^2\right),
\label{AAS2}
\end{eqnarray}
and the constraint equation becomes
\begin{equation} 
\sum_{i=1}^{n}x_i^2 \pm y^2 + z^2=1.
\end{equation}
Critical points correspond to fixed points where $x'_i=0$, $y'=0$ and
$z'=0$, and there are self-similar solutions with
\begin{equation} 
\frac{\dot{H}}{H^2}=-3\sum_{i=1}^{n}x_i^2 - \frac{3\gamma}{2}z^2. 
\end{equation}
This corresponds to an expanding universe with a scale factor $a(t)$ given
by $a\propto t^p$, where 
\begin{equation}
p=\frac{2}{6\sum_{i=1}^{n}x_i^2 + 3\gamma z^2}\,.
\end{equation}
The system (\ref{AAS1})-(\ref{AAS2}) has at most one $n$-dimensional 
sphere $S$ embedded in $(n+2)$-dimensional phase-space corresponding to 
kinetic-dominated solutions, a fixed point $A$ which is a
kinetic-potential-scaling solution, a fixed point $B$ which is a
fluid-dominated solution, and a fixed point $C$ which is a
fluid-potential-kinetic-scaling solution listed in Table 2.

$S$: $\sum_{i=1}^{n}x_i^2=1$, $y=0$, $z=0$. These kinetic-dominated
solutions always exist for any form of the potential, which are equivalent
to stiff-fluid dominated evolution with $a\propto t^{1/3}$ irrespective
of the nature of the potential. The linearization of system
(\ref{AAS1})-(\ref{AAS2}) about these fixed points yields
\begin{eqnarray*}
\delta x_i' &=& -2\left(\lambda_i \sqrt{\frac{3}{2}}-3x_i\right)
\sum_{j=1}^{n}(x_j \delta x_j), \\
\delta z' &=& \frac{3}{2}(2-\gamma)\delta z,
\end{eqnarray*}
which indicate that the solutions are marginally stable for
$\sum_{i=1}^{n}(\lambda_i x_i)>\sqrt{6}$ and a stiff fluid $\gamma=2$.

$A$: $x_i=\frac{\lambda _i}{\sqrt{6}}$,
$y=\sqrt{\pm(1-\frac{1}{6}\sum_{i=1}^{n} \lambda_i^2)}$, $z=0$.
The potential-kinetic-scaling solution exists for sufficiently flat
$\sum_{i=1}^{n}\lambda _i^2 < 6$ positive potentials or steep
$\sum_{i=1}^n\lambda _i^2 > 6$ negative potentials. The power-law exponent,
$p=\frac{2}{\sum_{i=1}^n \lambda_i^2}$, depends on the slope of the
potential. The linearization of system (\ref{AAS1})-(\ref{AAS2}) about this
critical point yields $(n+1)$ eigenvalues
\begin{eqnarray*}
m_i &=& -\frac{1}{2}\left(6-\sum_{j=1}^{n}\lambda_j^2\right), \\ 
m_z &=& -\frac{1}{2}\left(3\gamma-\sum_{j=1}^{n}\lambda_j^2\right).
\end{eqnarray*}
Thus the scaling solution is stable for a positive potential with
$\sum_{i=1}^{n}\lambda_i^2 < 3\gamma$, which corresponds to the power-law
inflation in an expanding universe when $\sum_{i=1}^n \lambda_i^2 < 2$.

$B$: $x_i=0$, $y=0$, $z=1$. The fluid-dominated solution exists for any
form of the potential, corresponding to a power-law solution with
$p=2/3\gamma$.
\begin{eqnarray*}
\delta x_i' &=& -3\delta x_i+(3\gamma-\sqrt{6}\lambda_i)\delta z, \\
\delta z' &=& 3 \gamma \delta z,
\end{eqnarray*}
which indicate that the solution is never stable.

$C$: $x_i=\sqrt{\frac{3}{2}}\frac{\gamma
\lambda_i}{\sum_{j=1}^{n}\lambda_j^2}$,
$y=\sqrt{\frac{3}{2}\frac{(2-\gamma)\gamma}{\sum_{i=1}^{n}\lambda_i^2}}$,
$z=\sqrt{1-\frac{3\gamma}{\sum_{i=1}^{n}\lambda_i^2}}$. The
fluid-potential-kinetic-scaling solution exists for a positive potential
with $\sum_{i=1}^{n}\lambda_i^2 > 3\gamma$. The power-law exponent,
$p=2/3\gamma$, is identical to that of the fluid-dominated solution,
depends only on the barotropic index $\gamma$ and is independent of the
slope $\lambda_i$ of the potential. The linearization of system
(\ref{AAS1})-(\ref{AAS2}) about the fixed point yields
\begin{eqnarray*}
\delta x_i' &=& 3(2-\gamma)x_i\sum_{j=1}^{n}(x_j\delta x_j)
-3\left( (1-\frac{\gamma}{2})(1-\sum_{j=1}^{n}x_j^2)
+\frac{\gamma}{2}y^2 \right)\delta x_i \\
 & & +(\sqrt{6}\lambda_i-3\gamma x_i)y \delta y,  \\
\delta y' &=& 3(2-\gamma)y\sum_{j=1}^{n}(x_j \delta x_j)
-\sqrt{\frac{3}{2}}\,y\sum_{j=1}^{n}(\lambda_j \delta x_j) \\
 & & +\left(\frac{3}{2}(2-\gamma)\sum_{j=1}^{n}x_j^2
-\sqrt{\frac{3}{2}}\sum_{j=1}^{n}(\lambda_j x_j)
+\frac{3\gamma}{2}-\frac{9\gamma}{2}y^2 \right)\delta y,
\end{eqnarray*}
which yield (n+1) eigenvalues
\begin{eqnarray*}
m_1 &=& -\frac{3(2-\gamma)}{4}\left(1+\sqrt{1-
\frac{8\gamma(\sum_{i=1}^{n}\lambda_i^2-3\gamma)}
{\sum_{i=1}^{n}\lambda_i^2(2-\gamma)}}\right), \\
m_2 &=& -\frac{3(2-\gamma)}{4}\left(1-\sqrt{1-
\frac{8\gamma(\sum_{i=1}^{n}\lambda_i^2-3\gamma)}
{\sum_{i=1}^{n}\lambda_i^2(2-\gamma)}}\right), \\
m_3 &=& \cdots = m_{n+1} = -\frac{3}{2}(2-\gamma).
\end{eqnarray*}
Thus the scaling solution is stable for a positive potential with
$\sum_{i=1}^{n}\lambda_i^2 > 3\gamma$.

The different regions in the ($\gamma,\lambda_i$) parameter space lead to
different qualitative evolution. For the sufficiently flat
($\sum_{i=1}^{n}\lambda_i^2 < 3\gamma$)
positive potential, $S$, $A$ and $B$ exist. Point $A$ is the stable
late-time attractor. Hence generic solutions begin in a kinetic-dominated
regime or at the fluid-dominated solution and approach the 
kinetic-potential-scaling solution at late times. For the intermediate
($3\gamma < \sum_{i=1}^{n}\lambda_i^2 < 6$) positive potential, all critical
points exist. Point $C$ is the stable late-time attractor. Hence generic
solutions start in a kinetic-dominated regime, at the
kinetic-potential-scaling solution or at the fluid-dominated solution and
approach the stable fluid-kinetic-potential-scaling solution. For the steep
($\sum_{i=1}^{n}\lambda_i^2 > 6$) positive potential, $S$, $B$ and $C$
exist. Point $C$ is the stable late-time attractor. Hence generic solutions
start in a kinetic-dominated regime or at the fluid-dominated solution and
approach the stable fluid-kinetic-potential-scaling solution. For the
sufficiently flat ($\sum_{i=1}^{n}\lambda_i^2 < 3\gamma$) negative
potential, the kinetic-dominated solution $S$ and the fluid-dominated
solution $B$ exist, which are unstable. For the intermediate
($3\gamma < \sum_{i=1}^{n}\lambda_i^2 < 6$) negative potential, the
kinetic-dominated solution $S$ and the fluid-dominated solution $B$ exist,
which are unstable. For the steep ($\sum_{i=1}^{n}\lambda_i^2 > 6$) negative
potential, $S$, $A$ and $B$ exist. Point $A$ is the stable late-time
attractor. Hence generic solutions start in a kinetic-dominated regime or
at the fluid-dominated solution and approach the stable
kinetic-potential-scaling solution at late times.

\begin{table}

\begin{tabular}{||c|c|c|c|c|c||} \hline \hline 
Label & $x_i$ & $y$ & $z$ & Existence & Stability \\ \hline
S & $\sum_{i}^{n}x_i^2=1$ & 0 & 0 & all $\lambda_i, \gamma$ &
stable $\gamma=2$,  \\
 & & & & & $\sum_{i}^{n}(\lambda_i x_i) > \sqrt{6}$ \\ \hline
A$_+$ & $\frac{\lambda_i}{\sqrt{6}}$ &
$\sqrt{(1-\sum_{i}^{n}\frac{\lambda _i^2}{6})}$ & 0 &
$\sum_{i}^{n} \lambda _i^2 < 6$ &
stable ($\sum_{i}^{n}\lambda_i^2 < 3\gamma$) \\ \hline 
A$_-$ & $\frac{\lambda_i}{\sqrt{6}}$ &
$\sqrt{(\sum_{i}^{n}\frac{\lambda _i^2}{6}-1)}$ & 0 &
$\sum_{i}^{n} \lambda _i^2 > 6$ & unstable ($V<0$) \\ \hline
B & 0 & 0 & 1 & all $\lambda_i, \gamma$ & unstable \\ \hline
C & $\sqrt{\frac{3}{2}}\frac{\gamma \lambda_i}{\sum_{j}^{n}\lambda_j^2}$ &
$\sqrt{\frac{3}{2}\frac{(2-\gamma)\gamma}{\sum_{i}^{n}\lambda_i^2}}$ &
$\sqrt{1-\frac{3\gamma}{\sum_{i}^{n}\lambda_i^2}}$ &
$\sum_{i}^{n}\lambda_i^2 > 3\gamma$ & stable ($V>0$) \\ \hline \hline
\end{tabular}

\caption{The properties of the critical points in a spatially flat FRW
universe containing $n$ scalar fields with the cross-coupling exponential
potential plus a barotropic fluid.}

\end{table}

\section{Conclusions and Discussions}

We have presented a phase-space analysis of the evolution for a spatially
flat FRW universe containing $n$ scalar fields with a positive or negative
cross-coupling exponential potential. In particular, for the
$\lambda _i=\lambda$ case, we find that in the expanding universe model with
a sufficiently
flat ($\lambda ^2 < 6/n$) positive cross-coupling potential the only
power-law kinetic-potential-scaling solution is the late-time attractor.
It is more difficult to obtain assisted inflation in such models since the
fields with cross-coupling exponential potential tend to conspire to act
against one another rather than assist each other. However, steep
($\lambda ^2 > 6/n$) negative cross-coupling potential has
kinetic-dominated solutions with $a \propto t^{1/3}$, some of which are 
the late-time attractors. It can be known that the kinetic energy of each
field tends to be equal via their effect on the expansion at late times.

Then we have extended the phase-space analysis of the evolution to a
realistic universe model with a barotropic fluid plus $n$ scalar fields
with a positive or negative cross-coupling exponential potential. We 
have shown that for the sufficiently
flat ($\sum_{i=1}^{n}\lambda_i^2 < 3\gamma$)
positive cross-coupling potential, the kinetic-potential-scaling solution
is the stable late-time attractor. The energy density of the scalar fields
dominates at late times. Moreover, for the steep
($\sum_{i=1}^{n}\lambda_i^2 > 6$) positive cross-coupling potential, the
fluid-kinetic-potential-scaling solution is the stable late-time attractor. 
However, a negative cross-coupling potential has no stable scaling
solutions.

\section*{Acknowledgements}

This project was in part supported by NNSFC under Grant
Nos. 10175070 and 10047004 as well as by NKBRSF G19990754.

\end{document}